\documentstyle[preprint,aps]{revtex}
\tighten
\begin{document}
\draft
\hbox to 16.5truecm{\hfil ROMA Preprint n. 1072-1994}
\hbox to 16.5truecm{\hfil hep-ph/9411341}

\vspace {2.0  cm}

\section*{\centerline {The neutrino cross section and upward going  muons} }

\vspace{\baselineskip}
\vspace{\baselineskip}
\centerline {Paolo Lipari, Maurizio Lusignoli and  Francesca Sartogo}

\vspace{\baselineskip}
\vspace{\baselineskip}
\centerline {\it Dipartimento di Fisica, Universit\`a di Roma ``la Sapienza",}
\centerline {\it and I.N.F.N., Sezione di Roma, Piazzale A. Moro 2,}
\centerline {\it I-00185 Roma, Italy}

\vspace {0.5 cm}

\begin{abstract}
The charged current cross section for neutrinos with energy of a few GeV
is reanalysed. In this  energy range the  cross section for the lowest
multiplicity exclusive channels is an  important fraction of $\sigma_{CC}$
and the approximation of describing the cross  section with deep inelastic
scattering formulae may be inaccurate. Possible consequences of our
reanalysis of the cross section in the interpretation of the data obtained
by deep underground detectors  on $\nu$--induced upward going  muons (both
stopping and passing)
are discussed.
\end{abstract}
\vspace{\baselineskip}
It is well  known that  measurements of the atmospheric neutrino fluxes
allow to perform sensitive studies on neutrino oscillations.
In some cases an anomaly has been observed, and interpreted as
positive evidence for neutrino oscillations \cite
{Kam_IMB_Soudan-contained,Kamioka-semi}.
Other experiments have
instead obtained results compatible with the no--oscillation
hypothesis, and gave therefore exclusion plots
in the neutrino oscillation parameter space ($\Delta m^2,\sin^2 2 \theta$)
\cite{No_oscill,IMB-up}.
In these studies it is often
necessary to compare experimental results  with theoretical
calculations that depend on assumptions about the neutrino fluxes
and interactions.

In this letter we  want to reanalyse the charged  current (CC)
neutrino and antineutrino cross sections with particular
attention  to the energy range  $E_\nu \lesssim 10$~GeV, and discuss
possible consequences for the interpretation of the measurements
of the atmospheric neutrino fluxes obtained by
deep underground detectors.
We  will argue  that it is possible to
improve  on the description of the cross sections
used in several recent analyses of neutrino-induced upward going muons
\cite{IMB-up,Eugeni,Frati}
including a more careful treatment of  the lowest
multiplicity channels (quasi--elastic scattering and single pion
production). The  main effect of this  more careful description
of the neutrino cross sections is an increase  of the  flux of low energy
upward going  muons.
Information on the energy spectrum of upward going muons can be obtained
from the measurement of the ratio of event numbers with muons stopping
in the apparatus or passing through \cite{IMB-up}.
Since with respect to previous calculations \cite{Eugeni,Frati}
we obtain a larger rate of stopping muon events
and the predicted rate
of passing events is increased by a smaller amount,
the excluded region in the neutrino oscillation parameter space
obtained from the measured ratio
stopping/passing could be significantly reduced.
The interpretation
of integral measurements  of the muon flux
with a low ($E_{\mu}^{\rm min} \sim 1$~GeV) energy threshold
could also be modified.

The flux of upward going muons of energy $E_\mu \ge E_\mu^{\rm min}$
and direction $\Omega$ can be calculated as
\begin{equation}
\Phi_{\mu^\mp} ( E_{\mu}^{\rm min}, \Omega) =
\int_{E_\mu^{\rm min}}^{\infty} dE_\nu
{}~\phi_{\nu_\mu(\overline {\nu}_\mu)} (E_\nu, \Omega)
{}~n_{\nu_\mu(\overline {\nu}_\mu) \to \mu^\mp}( E_\mu^{\rm min};E_\nu)\;,
\label{upw-flux-int}
\end{equation}
where $\phi_{\nu_\mu(\overline {\nu}_\mu)}(E_\nu, \Omega)$ is the
differential flux of $\nu_\mu$
($\overline {\nu}_\mu$) and
$n_{\nu_\mu(\overline {\nu}_\mu) \to \mu^\mp}(E_\mu^{\rm min};E_\nu)$ is
the average number  of muons above a threshold energy $E_\mu^{\rm min}$
produced  by a neutrino of energy $E_\nu$. It is given by
\begin{equation}
n_{\nu_\mu (\overline {\nu}_\mu) \to \mu^\mp} ( E_{\mu}^{\rm min};E_\nu) =
N_A \int_{E_{\mu}^{\rm min}}^{E_\nu} dE_0
{}~{d \sigma_{\nu_\mu(\overline {\nu}_\mu)} \over dE_0} (E_0, E_\nu)
{}~[R(E_0) - R(E_{\mu}^{\rm min}) ]\;.
\label{integral-yield}
\end{equation}
In equation (\ref{integral-yield}), $N_A$ is Avogadro's number and
the cross section refers to neutrino-nucleon CC scattering.
The energy $E_0$  of the muon  at the production point is
weighted  by a factor $d\sigma/dE_0$, the relevant cross section, and by a
factor $R(E_0) - R(E_\mu^{\rm min})$
($R(E)$ is the range in  rock of a muon  of energy $E$)
that takes into account the larger effective target available
with increasing muon energy.
Note that $n_{\nu_\mu (\overline {\nu}_\mu) \to \mu^\mp}$
depends not only on the total CC cross--section
but also on the shape  of the  muon energy spectrum.
In equation (\ref{upw-flux-int}) we are implicitly assuming
that the  observed muons are collinear with the parent neutrinos
($\Omega_\mu \simeq \Omega_\nu$) and in eq.(\ref{integral-yield})
we are neglecting fluctuations in the muon energy losses \cite{direction}.

The calculated muon flux (eq.\ref{upw-flux-int}) depends on
the inclusive cross-section for muon production.
In the literature \cite{IMB-up,Eugeni,Frati}
this cross section has been evaluated using the deep inelastic
scattering formalism.
The deep inelastic scattering (DIS) cross section,
expressed in terms of the
usual  kinematical variables
$y = 1 - E_\mu/E_\nu$ and $x = Q^2/(2 m_N E_\nu y)$,
must  be integrated over $x$ and $y$ up to the kinematical limits
which,
neglecting the lepton mass \cite{Jarlskog},
are $0\le x \le 1$ and
$0 \le y \le (1 + x\,m_N/(2\,E_\nu))^{-1}$.
We will denote this method of calculation as Method I.

We observe that the DIS formulae are expected to be valid
only for $Q^2$ sufficiently large, and that using them for calculating
the cross  section of low energy neutrinos implies an extrapolation
into a region  where nonperturbative effects may become important.
Most of the parametrizations of the parton distribution functions (PDF)
\cite {DIS_bigQ2} are valid only above a minimum
$Q^2_\circ$ of  several GeV$^2$.
The parton distributions for $Q^2 \le Q^2_\circ$
have been considered to be the same
as at $Q^2_\circ$, neglecting further evolution \cite{Eugeni,Frati}.
More recently new sets of parton distributions
that consider the evolution  down
to a lower value $Q^2_\circ \simeq 0.3$~GeV$^2$ have been made available
\cite{DIS_smalQ2}.

As a possible improvement we suggest to
consider separately the contributions of the exclusive channels
of lowest multiplicity, i.e. quasi--elastic  scattering and single pion
production, and  describe the additional channels  collectively
using the DIS formulae (Method II).
We decompose therefore
the CC neutrino  cross section as
the sum  of three contributions:
\begin{equation}
\sigma_{\nu(\overline {\nu})}^{CC}  = \sigma_{QEL} + \sigma_{1\pi} +
\sigma_{DIS} \;.
\label {sigma-decomp}
\end{equation}

The quasi--elastic cross section in eq.\ref{sigma-decomp}
is calculated following \cite{Llewellyn}.
The main uncertainty is in the axial--vector
form factor, for which we follow \cite{Belikov} assuming
$F_A (Q^2) = -1.25\,(1+Q^2 / M_A^2 )^{-2}$,
with $M_A$ = 1.0 GeV. Inclusion of nuclear effects would decrease
$\sigma_{QEL}$ by  $\sim 5\%$.

To determine the energy distribution of the muon
produced together with a single pion we make the
simplifying assumption of dominance of $\Delta(1232)$
production. The cross section is normalized to the results
of the more complete calculation by Fogli and Nardulli \cite{Fogli_Nar},
for a maximum mass of the pion-nucleon system $W_c =1.4\;{\rm GeV}$.
To avoid double counting, the deep inelastic scattering contribution
$\sigma_{DIS}$ is limited to the kinematic region where the mass of the
hadronic system in the final state  is $W \ge W_c$
 \cite{twopionprod}.
This corresponds  in the $(x,y)$ plane to the condition
$2\, m_N\, E_\nu\, y(1-x) \le W_c^2 - m_N^2$.

We observe that for neutrino energies
not much larger than the nucleon mass,  a large fraction
of the phase space corresponds to
$W \le W_c$. The deep--inelastic formula  in this region
does not take into account the detailed features of
the dynamics, with consequences both on the
absolute  value of the cross section and on the shape on the muon
spectrum. Moreover,
the kinematical region $m_N^2 <
W^2 < (m_N + m_\pi)^2$ is  unphysical. When it is included
in $\sigma_{DIS}$, as in Method I, it takes
(but only roughly) into account the quasi--elastic contribution.

The neutrino CC cross section  and its components calculated according to
equation (\ref{sigma-decomp}), using the PDF of Owens for $\sigma _{DIS}$,
are plotted in  fig.1 and compared with large statistics data
for the inclusive process
at high \cite{CCFRR} and low \cite{7feet} energy, and  with
data on one--pion production \cite{Onepionprod}
and quasi--elastic scattering \cite{Quasiel_exp}.
The exclusive channels contribute 89\% (12\%) of the cross section
for $E_\nu = 1$ (10) GeV.
Our model is explicitely non--scaling, the ratio $\sigma_\nu/E_\nu$ is not
a constant. In the  energy range
$E_\nu \sim 1$--10~GeV the ratio $\sigma_\nu/E_\nu$
is $\sim 20\%$ higher than the value measured at higher energy. At lower
energies,  closer to the threshold for muon
production,  $\sigma_\nu/E_\nu$  drops to zero.
The comparison with data \cite{PDG_92} of the neutrino CC cross sections is
rather encouraging, notwithstanding the large experimental errors.
The situation is  much less satisfactory for antineutrinos,
in that our recipe yields results systematically larger than
the sparse available data \cite{GGM_NUBAR}.
Note that the contribution of antineutrinos to the
total flux of positive and negative muons
is $\sim 1/3$ of the total.

We have calculated the flux of upward going muons
according to equation (\ref{upw-flux-int}) and its differential
version \cite{Eugeni}
using different  models of the neutrino fluxes
\cite{nu_fluxes}
and different parametrizations \cite{DIS_bigQ2,DIS_smalQ2}
for the nucleon structure functions
in $\sigma_{DIS}$.
In fig.2 we show the result of a differential muon flux  calculation,
in which the neutrino flux from Butkevich et al. \cite{nu_fluxes},
and the parton distributions given by Owens \cite{DIS_bigQ2}
have been  used.
The solid curve is obtained using
Method II for the cross section
(the dotted curve is the partial contribution of $\sigma_{DIS}$),
while the dashed curve  is  obtained
with Method I
(with results in excellent
agreement with other authors  \cite {Frati}).
The fluxes obtained with the two methods
are essentially equal for $E_\mu \gtrsim 10$~GeV,
but at lower muon energies Method II gives a
result that is considerably larger, by $\sim 12\%$ for $E_\mu = 3$~GeV and
by $\sim 25\%$ for $E_\mu = 1$~GeV. The contribution
of the exclusive channels in Method II amounts to $\sim 60\%\;(32\%)$
for $E_\mu = 1$~GeV (3 GeV).

We now discuss possible consequences of this
correction to the muon flux estimate.
The IMB collaboration  \cite{IMB-up} has  measured  a ratio of
rates of  stopping and passing upward going muons
 $(N_s/N_p) = 0.16 \pm 0.019$.
They have compared this  result with
a detailed  Montecarlo calculation
based on Method I, using the structure functions of EHLQ
\cite{DIS_bigQ2}
and the neutrino flux of Volkova, corrected at low energy
with the results of Lee and Koh \cite {nu_fluxes}, obtaining $0.163$ for
the stopping/passing ratio in the absence of oscillations.
It has been noted that in performing the ratio
stopping/passing the uncertainty due to
the absolute normalization of the neutrino flux
is greatly reduced and this has been verified in independent
calculations \cite{Eugeni,Frati}.

To  calculate a prediction for the rates of stopping and passing
upward going muons in a specific detector,
one needs of course a detailed
knowledge of acceptances and detection efficiencies.
As an approximation to the  real experimental situation,
following \cite{Eugeni}, we define
stopping muons those  in the  energy interval  $1.25 \le E_\mu \le 2.5$~GeV
and passing  muons those with $E_\mu \ge 2.5$~GeV,
assuming moreover that the detector acceptance and efficiency
for muons above $1.25$~GeV are approximately independent from
the direction and energy of the particles.
The fluxes $\Phi_{s}(\Phi_{p})$ are obtained integrating
eq.~\ref{upw-flux-int},
with the appropriate $E_{\rm min}$, on the entire downward hemisphere.

We report in Table I the results of calculations of $\Phi_{s}$,
$\Phi_{p}$ and their ratio obtained with different models for
the neutrino fluxes and different sets of leading--order
parton distributions. Both Method I (only DIS) and Method II
-- our preferred one -- have been used.
We observe that the results of Method II
are larger than those of Method I. For those parton
distributions that have a high $Q^2_\circ$ the
ratio $\Phi_{s}/\Phi_{p}$ is increased by $\sim 10\%$.
The variation is much less for parton distributions having low $Q^2_\circ$,
but it is interesting to note that Method II calculations using
different structure functions are in better agreement with each other.
The neutrino  fluxes of Volkova and Butkevitch (and Mitsui too)
predict  very similar $\Phi_s/\Phi_p$,
notwithstanding the differences of $\sim 10\%$ in normalization,
while the Bartol flux is flatter and predicts a ratio somewhat
smaller (by $\sim  5\%$).

In order to discuss the possible effects of neutrino oscillations,
we define the quantity
\begin{equation}
r_s = {(\Phi_s/\Phi_p) \over(\Phi_s/\Phi_p)_\circ}\;,
\end{equation}
where $(\Phi_s/\Phi_p)_\circ$ is the stopping/passing ratio
in absence of neutrino oscillations.
The  ratio $r_s$  depends on
the neutrino oscillation parameters ($\Delta m^2,\sin^2 2 \theta$).
We will only consider here the case of $\nu_\mu \leftrightarrow
\nu_\tau$ oscillations.

To take into account the  effects of oscillations we have to  make
in eq. (\ref{upw-flux-int})
the substitution
$\phi_{\nu_\mu} \to [1 - P_{\nu_\mu \to \nu_\tau}]~ \phi_{\nu_\mu}$,
and analogously for antineutrinos.
The double ratio $r_s$ may take the values
$r_{\rm min} \le r_s \le 1$.
With our definitions of $\Phi_s$ and $\Phi_p$,
the maximum  suppression of the stopping to passing ratio is
$r_{\rm min} \simeq 0.66$,
corresponding to
$\sin^2 2 \theta = 1$ and $\Delta m^2 \simeq 1.3 \times 10^{-3}$~eV$^2$.
The ratio $r_s$  becomes  unity in the limit
$\sin^2 2 \theta \to 0$ and/or $\Delta m^2 \to 0$, that correspond to the
no oscillation case, and also for $\Delta m^2 \to \infty$. In this limit
(in practice for $\Delta m^2 \gtrsim 1$~eV$^2$)
neutrinos of all significant  energies
oscillate  many times, so that the spectrum is suppressed
without distortions by a constant factor $1 - {1\over 2} \sin^2 2 \theta$,
and the stopping/passing ratio remains unchanged.

In fig.3 we have drawn lines of constant $r_s$ in the
$\Delta m^2,\sin^2 2 \theta$ plane.
The region favoured at $90\%$ c.l. by the recent Kamiokande-III analysis
\cite{Kamioka-semi} is also shown, limited by the dashed line.
As one can see, parameters in this region
imply measurable effects in observations of upward going muons:
in fact the region to the right of the curve $r_s = 0.8$ corresponds
approximately to the region excluded by the (Method I) analysis of
the IMB collaboration \cite{IMB-up}.

Let us assume that $r_s^\ast$ be the ratio of an experimental
result with the theoretical no--oscillation value and its error
$ \Delta  r_s^\ast$  includes both the experimental error
and the  systematic  uncertainty in the theoretical
calculation. At $90\%$ c.l. one could exclude
in the ($\Delta m^2$, $\sin^2 2 \theta$) plane the region corresponding
to values $|r_s - r_s^\ast| \ge 1.64\,\Delta  r_s^\ast$
\cite{statist}. Use of Method II for the neutrino cross section increases the
theoretical prediction by $\sim 10\%$, therefore  $r_s^\ast$ would be lowered
by the same amount, with obvious consequences on the excluded region.
We urge therefore the experimental groups that have collected and are
collecting data on upward going muons to reanalyze them going beyond
the DIS approximation for the $\nu$ cross section.

P.L. acknowledges useful conversations with T.K. Gaisser and T. Stanev,
M.L. support by the European Community under HCM Programme, contract
CHRX-CT93-0132.

\newpage
\begin{table}
\caption{
Fluxes (in units $10^{-13}$~(cm$^2$\,s\,sr)$^{-1}$) of  `stopping'
($1.25 \le E_\mu \le 2.5$~GeV) and  `passing'
($E_\mu \ge 2.5$~GeV) upward going muons calculated
with different models  of the
neutrino fluxes, different choices of the parton distribution
functions   and using Method I and Method II to describe
the neutrino cross section.}
\vspace{0.3 cm}
\begin {tabular} { l  c  c  c  c  c  c  c  c  c  }
{}~~~ & \multicolumn {3} { c } {EHLQ--2} &
\multicolumn {3} { c } {Owens} &
\multicolumn {3} { c } {MRS--LO} \\
\hline
$\nu$--flux &
$\Phi_s$ & $\Phi_p$ & $\Phi_s/\Phi_p$  &
$\Phi_s$ & $\Phi_p$ & $\Phi_s/\Phi_p$  &
$\Phi_s$ & $\Phi_p$ & $\Phi_s/\Phi_p$ \\
\hline
 Volkova (II) &   0.453 &   2.140 &   0.212 &   0.484 &   2.361 &
 0.205 &   0.492  & 2.409  &  0.204\\
 Volkova  (I) &   0.383 &   2.053 &   0.187 &   0.432 &   2.295 &
 0.188 &   0.476  &  2.387  &  0.199\\
(II)/(I)       &    1.18 &    1.04 &    1.13 &    1.12 &    1.03 &
 1.09 &    1.03  &  1.01  &  1.03\\
\hline
 Butkevich (II)   &   0.507 &   2.378 &   0.213 &   0.544 &   2.622 &
 0.207 &   0.553  &  2.678  &  0.206 \\
 Butkevich (I)  &   0.433 &   2.275 &   0.190 &   0.488 &   2.544 &
 0.192 &   0.536  & 2.652  &  0.202 \\
(II)/(I)        &    1.17 &    1.05 &    1.12 &    1.11 &    1.03 &
 1.08 &    1.03  &  1.01  &  1.02  \\
\hline
 Bartol (II)  &   0.454 &   2.308 &   0.197 &   0.487 &   2.547 &
 0.191 &   0.496  &  2.601  &  0.191  \\
 Bartol (I)   &   0.389 &   2.214 &   0.176 &   0.439 &   2.475 &
 0.177 &   0.482  &  2.578  &  0.187\\
(II)/(I)        &    1.17 &    1.04 &    1.12 &    1.11 &    1.03 &
 1.08 &    1.03  &  1.01  &  1.02  \\
\end{tabular}
\end{table}

\vspace{\baselineskip}
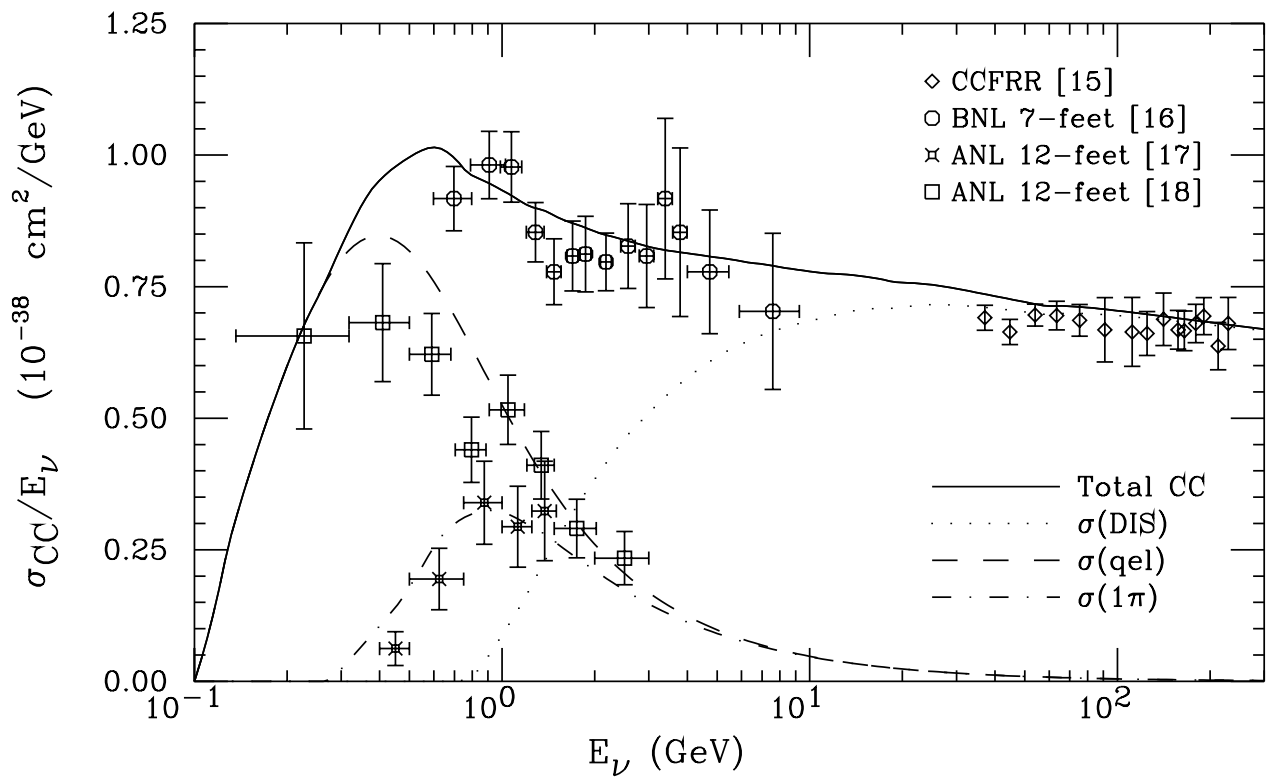
\begin{figure}
\caption{$\nu_\mu$ CC cross sections plotted as a function of energy.}
\label{1}
\end{figure}
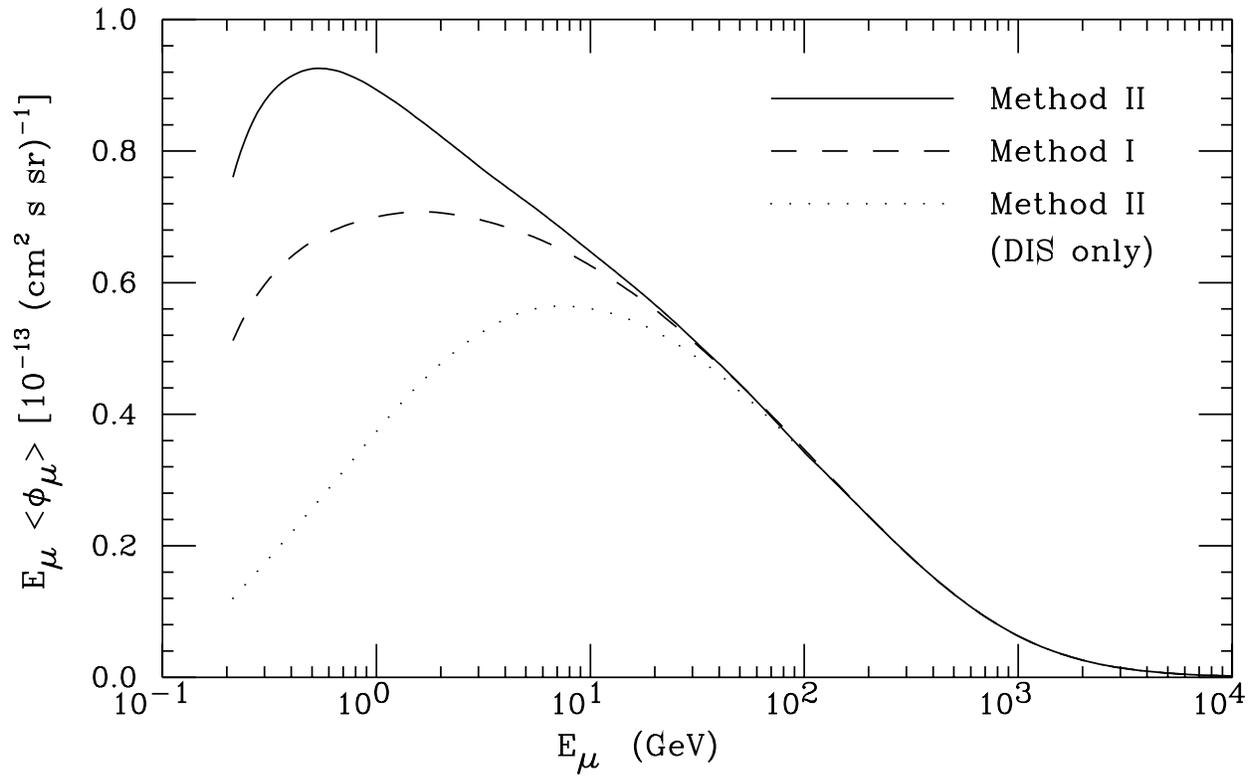
\begin{figure}
\caption{Differential flux of upward going muons averaged in angle over
one hemisphere and plotted as a function of muon energy.}
\label{2}
\end{figure}
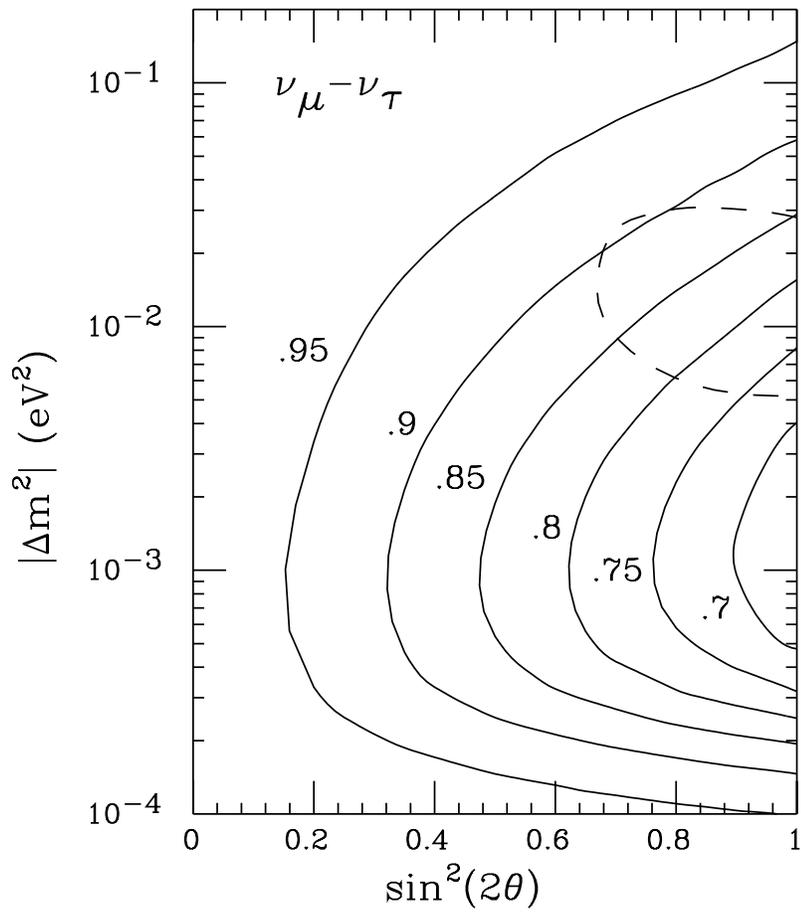
\begin{figure}
\caption{
Curves in the ($\Delta m^2$, $\sin^2 2 \theta)$  plane that correspond to
constant values  for the double ratio  $r_s$.
Also shown (dashed) is the $90\%$ c.l. curve obtained by the Kamiokande-III
combined analysis of contained and semi--contained events
\protect\cite{Kamioka-semi}.}
\label{3}
\end{figure}


\begin{thebibliography}{999}

\bibitem {Kam_IMB_Soudan-contained} K.S. Hirata  {\it et al.}
(Kamiokande-II Coll.), Phys.Lett.B {\bf 205}, 416 (1988), {\it ibid.}
 {\bf 280}, 146 (1992); D. Casper {\it et al.} (IMB Coll.),
Phys.Rev.Lett. {\bf 66}, 2561 (1991); M.C. Goodman {\it et al.}
(Soudan-2 Coll.) in Proc. 23rd  Int. Cosmic Ray Conf. (Calgary)
{\bf 4}, 446, (1993).

\bibitem {Kamioka-semi} Kamiokande-III Collaboration, paper presented
by M. Nakahata to ICHEP94 Conference, Glasgow, to appear in the Proceedings.

\bibitem {No_oscill} Ch. Berger {\it et al.} (Fr\'ejus Coll.),
Phys.Lett.B {\bf 245}, 305 (1990), {\it ibid.} {\bf 227}, 489 (1989);
M.Aglietta {\it et al.} (Nusex Coll.), Europhys.Lett.
{\bf 8}, 611 (1989);
M. Mori {\it et al.} (Kamiokande Coll.),
Phys.Lett.B {\bf 270}, 89 (1991);
M.M. Boliev {\it et al.} (Baksan Coll.),
in Proc. 3rd Int. Workshop on Neutrino Telescopes (M. Baldo Ceolin, ed.),
Venice, 1991, p.235.

\bibitem {IMB-up} R. Becker-Szendy {\it et al.} (IMB Coll.),
Phys.Rev.Lett. {\bf 69}, 1010 (1992).

\bibitem {Eugeni} E. Akhmedov, P. Lipari and M. Lusignoli,
Phys.Lett.B {\bf 300}, 128 (1993).

\bibitem{Frati}  W. Frati, T.K. Gaisser, A.K. Mann and T. Stanev,
Phys.Rev.D {\bf 48}, 1140 (1993).

\bibitem{direction} For a discussion of the collinearity
approximation, see \cite{Frati}. Muon energy loss
fluctuations are negligible here
(see  P. Lipari and T. Stanev, Phys.Rev.D {\bf 44}, 3543 (1991)).

\bibitem{Jarlskog} For nonzero lepton mass formulae, see
C.H. Albright and C. Jarlskog, Nucl.Phys.B {\bf 84}, 467 (1975).

\bibitem {DIS_bigQ2} E. Eichten, I. Hinchcliffe, K. Lane and C. Quigg
(EHLQ), Rev.Mod.Phys. {\bf 56}, 579 (1984) and {\bf 58}, 1065 (1986) (E);
J.F. Owens  Phys.Lett.B {\bf  266}, 126 (1991).

\bibitem{DIS_smalQ2} M. Gluck, E.Reya and A.Vogt,
Z.Phys.C {\bf 53}, 127 (1992);
A.D. Martin, W.J. Stirling and R.G. Roberts (MRS),
Phys.Rev.D {\bf 47}, 867 (1993) and preprint DTP/94/78 (hep-ph/9409410).

\bibitem {Llewellyn}  C.H. Llewellyn Smith, Phys.Rep. {\bf 3}, 271 (1971).

\bibitem {Belikov}  S.V. Belikov, Z.Phys.A {\bf 320}, 625 (1985).

\bibitem {Fogli_Nar} G.L. Fogli and G. Nardulli, Nucl.Phys.B {\bf 160},
116 (1979).

\bibitem{twopionprod} We have neglected the contribution of
two pion production with $W \le W_c$, which we expect to
be quite small.

\bibitem{CCFRR} D.B. McFarlane {\it et al.} (CCFRR), Z.Phys.C {\bf 26},1
(1984).

\bibitem{7feet}  N.J.Baker {\it et al.} Phys.Rev.D {\bf 25}, 617 (1982).
Reference to other low energy experiments can be found in
\protect\cite{PDG_92}.

\bibitem{Onepionprod}  S.J. Barish {\it et al.},
Phys.Rev.D {\bf 19}, 2521 (1979).

\bibitem{Quasiel_exp} S.J. Barish {\it et al.}, Phys.Rev.D {\bf 16},
3103 (1977).

\bibitem{PDG_92} Rev. Particle Properties, Phys.Rev.D {\bf 45}, part II
(1992).

\bibitem {GGM_NUBAR} O.Erriquez {\it et al.}, Phys.Lett. {\bf 80}B, 309 (1979).

\bibitem {nu_fluxes} L.V. Volkova, Yad.Fiz. {\bf 31}, 784 (1980)
(Sov.J.Nucl.Phys. {\bf 31}, 1510 (1980));
A.V. Butkevich, L.G. Dedenko and  I.M. Zheleznykh,
Yad.Fiz. {\bf 50}, 142 (1989) (Sov.J.Nucl.Phys. {\bf 50}, 90 (1989));
K. Mitsui, Y. Minorikawa and H. Komori, N.Cim.C {\bf 9}, 995 (1986);
V. Agrawal, T.K.Gaisser, P. Lipari and T. Stanev, paper in  preparation
(``Bartol''); H. Lee and Y.S. Koh, N.Cim.B {\bf 105}, 883 (1990).

\bibitem{statist}
This is true for gaussian errors and far from the boundaries of the
physical region (for a discussion, see \protect\cite{PDG_92}).
Note that assuming \protect\cite{Eugeni} a $12\%$ relative
error on experiment, a much smaller theoretical error,
and $r_s^\ast=1$, one obtains for the $90\%$ c.l.
lower limit $r_s \ge 0.8$.

\end{thebibliography}
\end{document}